# Experimental Search for the Singlet Metastable Deuteron in the Radiative *n-p* Capture.

S.B. Borzakov, N.A. Gundorin and Yu.N. Pokotilovski

Joint Institute for Nuclear Research
141980 Dubna, Moscow region, Russia

We performed an experimental search for the bound state singlet deuteron predicted in some microscopic calculations. The experiment consists in a high statistics measurement of $\gamma$-ray spectra after thermal neutron capture by hydrogen nuclei. The upper limit is obtained for the probability of the $^3S_1 \to {}^1S_0$ $\gamma$-transition population of the deuteron singlet bound state with the bound energy in the interval 25-125 keV.

1. Introduction.

Investigations of nucleon-nucleon interaction are very important for understanding nuclear forces. The neutron-proton scattering amplitude in the simplest effective range approach is [1]

$$F = \frac{1}{g(k) - ik}, \qquad (1)$$

where

$$g(k) = k \cdot ctg\,\delta = -\frac{1}{a} + \frac{1}{2}\rho k^2. \qquad (2)$$

Here $k$ – is the neutron momentum, $a$ is the neutron-proton scattering length, $\rho$ is the effective range.

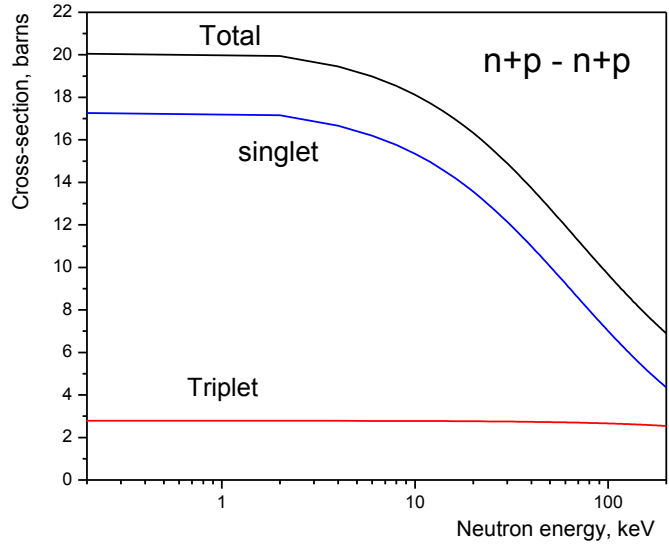

*Fig. 1. The calculated energy dependence of the **np** scattering cross-sections in different spin states.*

Parameters of this model for both singlet and triplet states of the **np** system are determined from the experimental spin-depended slow neutron-proton scattering cross-sections (see Table 1).

It is assumed that poles of the S-matrix

$$S = 1 + 2ikF(k) \qquad (3)$$

at negative energy of the **np** system ($k = i \cdot \kappa$) correspond to the bound triplet state (ground state of the deuteron, $\kappa = 0.232 Fm^{-1}$) and to the quasi-bound singlet one ($\kappa = -0.044 Fm^{-1}$, virtual level). It is widely believed that the singlet (**np**) system is not bound (see also [2]). Other poles as well as poles appearing in more precise presentation of the function *g(k)* [3]

$$g(k) = -\frac{1}{a} + \frac{1}{2}\rho k^2 + Pk^4 + Qk^6 + \ldots \qquad (4)$$

are considered to be unphysical.

Table 1. Scattering parameters.

| State | Scattering length $a$, Fm | Effective range $\rho$, Fm | $\kappa_{1,2}$, Fm$^{-1}$ | E, MeV |
|---|---|---|---|---|
| $np$ ($^1S_0$) | $-23.719 \pm 0.013$ | $2.76 \pm 0.05$ | $-0.044$ | $-0.080$ |
|  |  |  | $1.20$ | $-59.6$ |
| $np$ ($^3S_1$) | $5.414 \pm 0.005$ | $1.750 \pm 0.005$ | $0.232$ | $-2.225$ |
|  |  |  | $0.911$ | $-34.4$ |

Usually effective range theory describes the scattering only. The radiative **np** capture is described in the effective range approach with a complex scattering length $A = a - i \cdot b$ so that capture cross-section is $\sigma_c = \dfrac{\pi \cdot b}{k}$ at $E_n \to 0$ [4].

"Virtual levels" are not quite terminologically defined. For example in the monograph of Goldberger and Watson [5] the scattering resonances with positive energy are called "virtual states". Other publications (Landau and Lifshits [1], Bethe and Morrison [6], Baz', Zel'dovich and Perelomov [7]) considering particular deuteron singlet state do not determine the sign of its energy. Ma [8] and Hulten and Sugawara [9] determine deuteron singlet virtual state as having positive energy.

It was said in [10] that only in the absence of bound states which are able to explain specific behavior of the cross sections at low energy it is possible to declare the presence of the anti-bound (virtual) states.

Is it possible to use the idea of negative resonance instead of the virtual level [8, 11, 12] as a different phenomenological model of **np** scattering? The scattering amplitude can be presented as:

$$F = \dfrac{1}{-\dfrac{1}{a-ib} + \dfrac{1}{2}\rho k^2 - ik} = \dfrac{1}{2k}\dfrac{\dfrac{4}{\rho}k}{k^2 - \dfrac{2}{\rho a} - i\dfrac{1}{2}\left(\dfrac{4k}{\rho} + \dfrac{4b}{\rho a^2}\right)} \quad (5)$$

$$= \dfrac{1}{2k}\dfrac{\Gamma_n}{E - E_r - i\dfrac{1}{2}(\Gamma_n + \Gamma_\gamma)}$$

One can see that this formula has the Breit-Wigner view with the neutron width $\Gamma_n = \dfrac{4k}{\rho} \propto \sqrt{E_n}$ and the radiative width $\Gamma_\gamma = \dfrac{4b}{\rho a^2}$ (in units of $\dfrac{\hbar^2}{2\mu} = 41.47 MeV \cdot Fm^2$, $\mu$ is the nucleon mass). The resonance energy is $E_r = \dfrac{2}{ar}\dfrac{\hbar^2}{2\mu}$ and is negative for the singlet state. The continuation of the scattering amplitude at negative energy is possible in two ways:

the neutron width is zero or is imaginary. In the latter case the pole for scattering amplitude corresponds to the virtual level.

This model describes the scattering and the radiative capture as well. According to [11] the radiative resonance width is of the order 10 eV and it may be observed in the resonance gamma ray scattering or in the **np** capture. F.L. Shapiro et al. [13] used the approach of negative resonance to describe the capture reaction $n + {}^3He \rightarrow T + p$. It was confirmed in the proton – tritium scattering.

2.  Theoretical predictions.

There are theoretical predictions of the bound deuteron singlet state (metastable in respect to $\gamma$-ray deexcitation) what is in compliance with the idea of negative resonance. Ivanov et al. [14] considered the deuteron as the Cooper **np** pair in field theoretical approach within Nambu - Iona-Lasinio model of the light nuclei. For the Cooper pair in the ${}^1S_0$ state they computed the binding energy $B({}^1S_0) = 79 \pm 12$ keV. The calculations agree well with the energy of the virtual level defined from the experimental S-wave scattering length. Maltman and Isgur [15] described **np** system as six quark state and they have obtained next values for the binding energies: 400 ± 400 keV for the singlet state and 2.9 Mev for the triplet state. Lattice quantum chromodynamics calculations [16] predict the existence of the bound singlet deuteron. Hackenburg [17] employed in his calculations the intermediate off-shell singlet and triplet deuterons treated as dressed dibarions. In a simple extension of the effective range theory he predicted the existence of the spin-singlet deuteron bound state. The binding energy of this singlet level was predicted to be 66 keV. He showed that an account of the radiative capture leads to the possibility of observation of the metastable singlet level in resonance scattering of gamma quanta by deuterons and in the two-photon radiative capture with the expected cross-section 27 μb, four orders of magnitude less than main **np** radiative capture channel.

The **np** ${}^1S_0$ radiative capture cross-section may be calculated according to [18]:

$$\sigma_{n\gamma}(M1) = 2\pi\alpha \frac{c}{v_n} (\mu_n - \mu_p)^2 \left(\frac{B_d}{Mc^2}\right)^{5/2} (\gamma^{-1} - a_S)^2 \qquad (6)$$

$\alpha$ - is the fine structure constant, **c** is velocity of light, $v_n$ is the neutron velocity, $\mu_n$ and $\mu_p$ are the neutron and proton magnetic moments, $B_d$ – is the deuteron binding energy, **M** is the nucleon mass, $a_s$ is the singlet scattering length. At the thermal neutron velocity $2.2 \times 10^5$ cm/s, and $B_d = 2224$ keV we have $\sigma_{n\gamma}(M1)$ = 300 mb close to the experimental value $\sigma(n_{th} p \rightarrow d\gamma) = 334 \pm 0.5$ mb.

For the hypothetical resonance singlet level similar calculation of the radiative capture $^3S_1 \to {}^1S_0$ cross section gives the ratio of the two gammas to one gamma transition

$$\frac{\sigma_{2\gamma}}{\sigma_\gamma} = \frac{(\gamma_0 - a_t^{-1})^2}{(\gamma - a_s^{-1})^2}\left(\frac{a_t}{a_s}\right)^2\left(\frac{B_S}{B_d}\right)^{3/2} \quad (7)$$

if to replace for the first $^3S_1 \to {}^1S_0$ transition the singlet scattering length by the triplet one and the deuteron binding energy $B_d$ = 2224 keV by the hypothetical singlet binding energy $B_s$ = 67 keV ($\gamma_0 = \frac{1}{\hbar}\sqrt{2M_N B_s}$). The obtained in this way estimate agrees with [17].

3. Experimental indications.

There are some experimental demonstrations of existence of the singlet deuteron. Cohen et al. [19, 20] observed the singlet deuteron in reaction $^9Be(p,d_s)^8Be$ at the energy of incident protons 12 MeV. Gaiser et al. [21] investigated reaction $^4He(d,p\alpha)n$ at the energy of bombarding deuterons 7 MeV. Their data gave clear evidence for the production of the singlet deuterons. Bohne et al. observed the analogous process in the reaction ($^3He, d_s$) [22]. Bochkarev et al. [23] investigated decays of the excited $2^+$ states of the $^6$He, $^6$Li, $^6$Be nuclei. From the energy and momentum conservation the narrow peaks in the α - spectra were considered as indications of the two particles decays: α - particle and the singlet deuteron in the case of $^6$Li and α –particle and the dineutron in the case of $^6$He. Attempts to describe the experimental spectra in terms of the two-nucleon final state interaction lead to abnormally large nucleon-nucleon scattering lengths 50-100 F.

Generally the problem of existence of the singlet deuteron is closely connected to old problem of existence of dineutron [24] and more generally of the neutral nuclei. Experimental search for dineutron was the subject of a number of experiments. In some of them there were indications of observation of the dineutron [23, 25], the tetraneutron [26], and even multineutrons with number of neutrons $n \geq 6$ [27] and very recently octaneutron [28].

## 4. Experiment.

The experiment consisted in recording gamma ray pulse height spectra from thermal neutron capture by protons: $n_{th} + p \to d + 2\gamma$. Existence of the singlet deuteron could be evidenced by a two-step transition $^3S_1 \to {}^1S_0 \to {}^3S_1$ in addition to the direct 2223 keV $^1S_0 \to {}^3S_1$ one. In this experiment we search for a weak peak not belonging to gamma radiation from elements contained in surrounding materials. The main interest was concentrated in the energy range $\approx 100$ keV below the gamma line with $E_\gamma = 2223$ keV, but larger energy interval was analyzed as well.

This work is a continuation of the experiment [29] performed with the correlation gamma-spectrometer COCOS [30] at the JINR pulse reactor IBR-2 with a power of 2 MW and frequency 5 Hz in a direct thermal neutron flux of $5 \times 10^5$ n/cm$^2$/s. We investigated in that work the gamma-ray energy range (2134-2158) keV chosen on the base of the prediction of Ref. [17] and obtained the upper boundary for the cross section of the searched for two-step process $\sim 50 \mu b$.

This measurement was carried out at modernized pulse reactor IBR-2M. In the present experiment thermal neutrons with a mean energy of about 30 meV were conducted to the target area by the curved 15 m long neutron mirror guide. The distance from the neutron moderator to the target was about 23 m, including air gaps between the moderator and entrance to the guide, and from the guide exit to the target, where the beam was collimated to the desired size 1.5x4.5 cm$^2$ with a mean intensity of about $6 \times 10^5$ cm$^{-2}$s$^{-1}$. The fast neutron and gamma-ray background was significantly lower than in a preceding experiment but was still rather high.

The scheme of the experiment is shown in Fig.2. The measurements were performed with targets from polyethylene, graphite and without any target. The targets were placed in the thin-walled aluminum container diam. 22×30 cm, 5 mm thick filled with 90% enriched $^6$LiF, preventing scattered thermal neutrons from being captured in HPGe detector and surrounding materials. Three different size polyethylene targets were used: 1) diam. 15.5 cm ×21 cm with a hollow 3×8×14 cm for the neutron beam; 2) disc diam. 15.5×7 cm; 3) disc diam. 9.5×4.8 cm with a hollow diam. 5×4 cm. The data taking was broken into a series of separate runs with different targets. The measurements were carried out with the HpGe detector (ORTEC) with relative efficiency 35% and energy resolution about 2.1 keV for 1173 keV line of $^{60}$Co. The detector was surrounded by the lead shield and was placed at the distance (15-20) cm from the target. The energy calibration was determined from gamma-ray peaks from $^{60}$Co, $^{137}$Cs and other sources and from the first and second-escape peaks of 2223 keV line.

The pulsed structure of the incident thermal neutron beam (see Fig. 3) was another problem of the experiment: the maximal loading at the peak of the neutron time-of-flight spectrum exceeded the mean one many times. The maximal count rate of the spectrometer could not exceed $2.4 \cdot 10^4$ sec$^{-1}$. The characteristics of the count rate loading of the spectrometer are shown in Table 2.

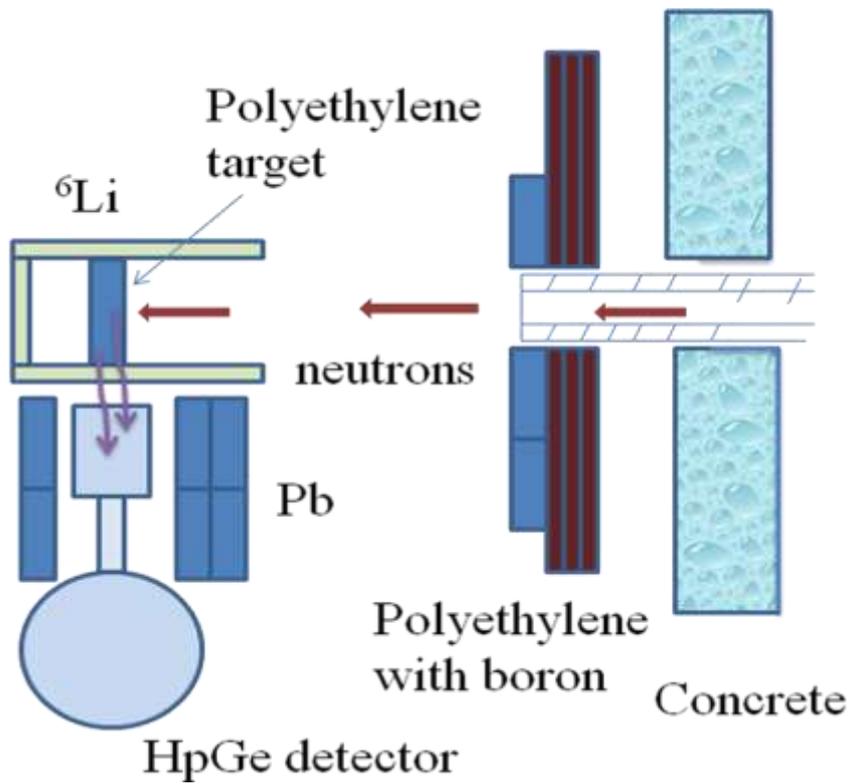

*Fig. 2. The scheme of the experiment.*

Table 2. Count rate loading of the spectrometer in measurements with different targets.

| Target | Total count rate, s$^{-1}$ | Counts in the energy interval, keV | | | R |
|---|---|---|---|---|---|
| | | 2171-2200 | 2201-2240 | 2241-2300 | |
| No. 1 (45 h) | 5.86·10$^3$ | 4.12·10$^6$ | 6.56·10$^7$ (376 s$^{-1}$) | 5.94·10$^6$ | 13.0 |
| No. 2 (52 h) | 1.47·10$^3$ | 6.98·10$^5$ | 1.71·10$^7$ (86 s$^{-1}$) | 1.53·10$^6$ | 15.3 |
| No.2 (100.5h) 1 disk, small hole | 1.37·10$^3$ | 1.30·10$^6$ | 3.27·10$^7$ (85.6 s$^{-1}$) | 2.62·10$^6$ | 16.7 |
| H1das4 (110 h) | 1.63·10$^3$ | 2.05·10$^6$ | 4.49·10$^7$ (106 s$^{-1}$) | 4.17·10$^6$ | 14.4 |
| No. 3 (211 h) | 2.24·10$^3$ | 5.64·10$^6$ | 1.23·10$^8$ (152 s$^{-1}$) | 1.16·10$^7$ | 14.3 |

The last column shows the ratio $R = \dfrac{2 \cdot N_{peak}}{(N_1 + N_2)}$, where $N_1$ and $N_2$ are the counts in the intervals left and right from the peak at 2223 keV. This value is a characteristics of the "effect"/"background" ratio in the measurements with different targets. The measurements with the target No. 1 gave too large overloading and spreading of the peaks and were not taken into account in the final analysis.

The background measured without target was measured in two ways: when the neutron beam passed to the remote wall of the experimental box and when the neutron beam was absorbed in the $^6$LiF shield. The background consists of three parts: the natural background at zero reactor power, background from the neutron guide, and from gamma rays generated from materials surrounding the target in the walls of the experimental box. The main difficulty of this experiment was caused by the latter component of the background.

The background was measured also with the 2 cm thick graphite target. Unfortunately this was not equivalent scatterer to the polyethylene target due to large difference in the scattering cross sections.

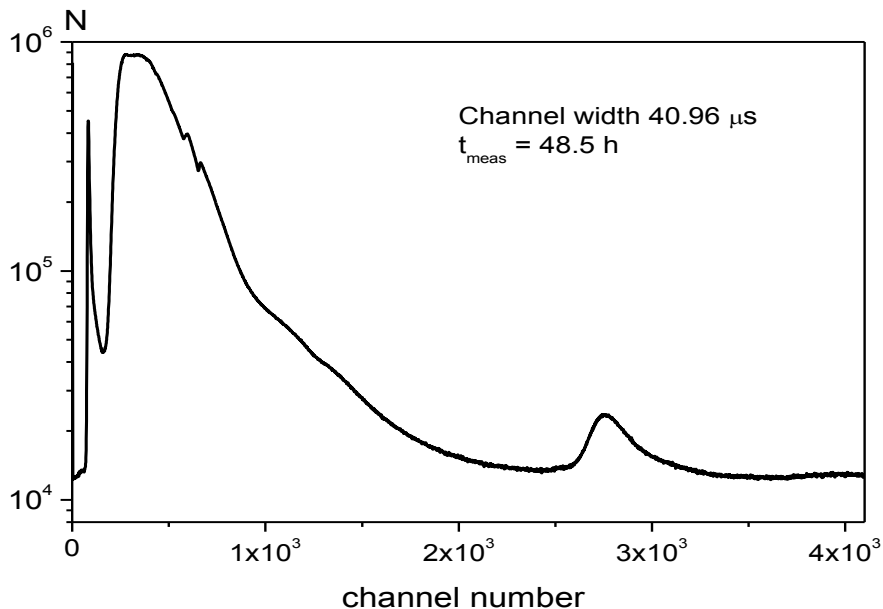

*Fig. 3 Time of flight spectrum, measured with the target No. 1.*

5. Experimental results.

Fig. 4 shows part of the pulse height spectrum, compiled from measurements totally ranging 532 hours with two polyethylene targets.
Analysis of the spectra has been performed with program VACTIV [31] allowing determination of peaks, their energy and peaks areas. Numerous peaks from neutron capture by $^{56}$Fe, $^{27}$Al, $^{12}$C, $^{14}$N etc. were determined. Identification of the peaks has been performed with use of tables [32]. There were discovered also gamma rays produced after neutron activation, for example from $^{116m}$In, $^{56}$Mn, $^{28}$Al etc. The yields and energies of these gamma quanta were determined with help of Tables in [33].

Next two figures: 5 and 6 show different parts of the gamma spectrum. Fig. 7 shows part of the spectrum in vicinity of the main transition 2223 keV line: energy interval 2100-2210 keV.

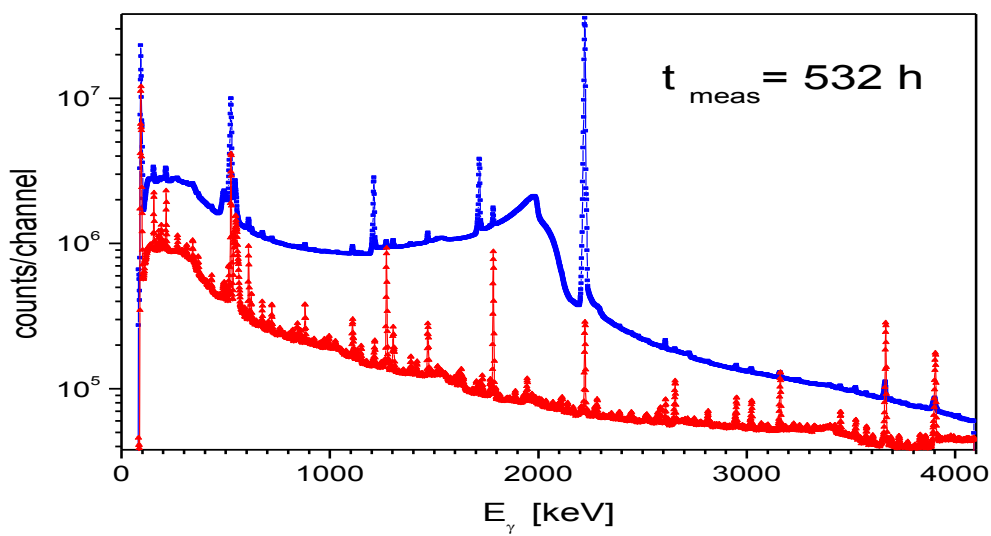

*Fig. 4. The gamma ray spectra measured with polyethylene target. Low curve shows the spectrum obtained with the graphite target.*

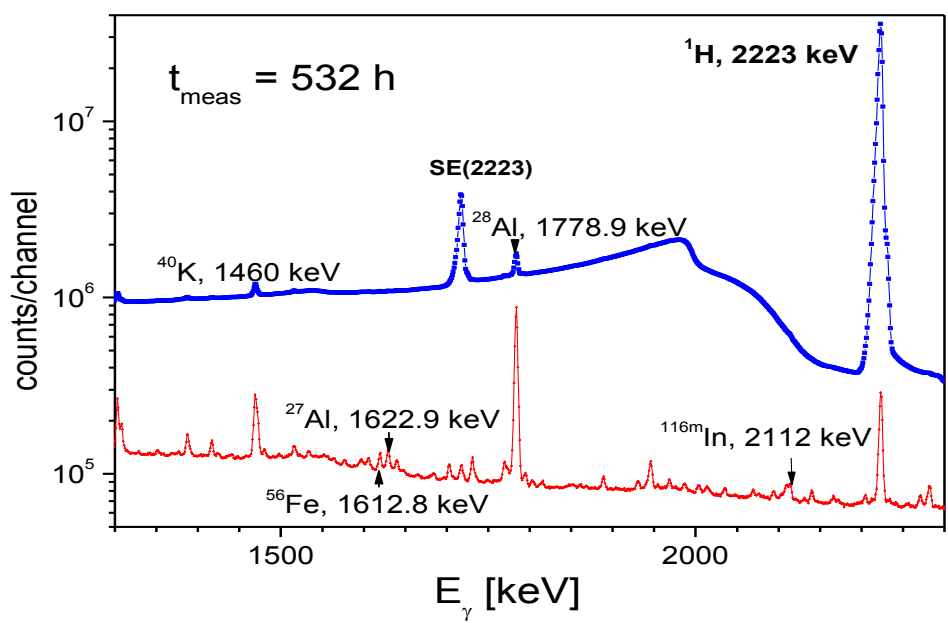

*Fig. 5. The part of the gamma ray spectra measured with polyethylene target.*

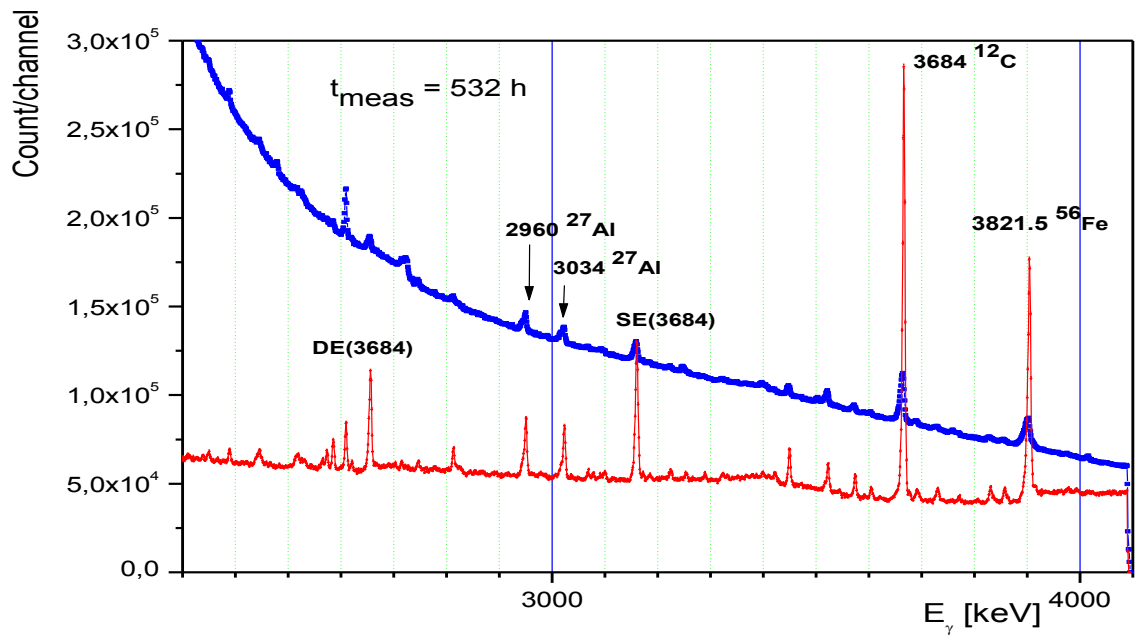

*Fig. 6. The part of the gamma ray spectra measured with polyethylene target.*

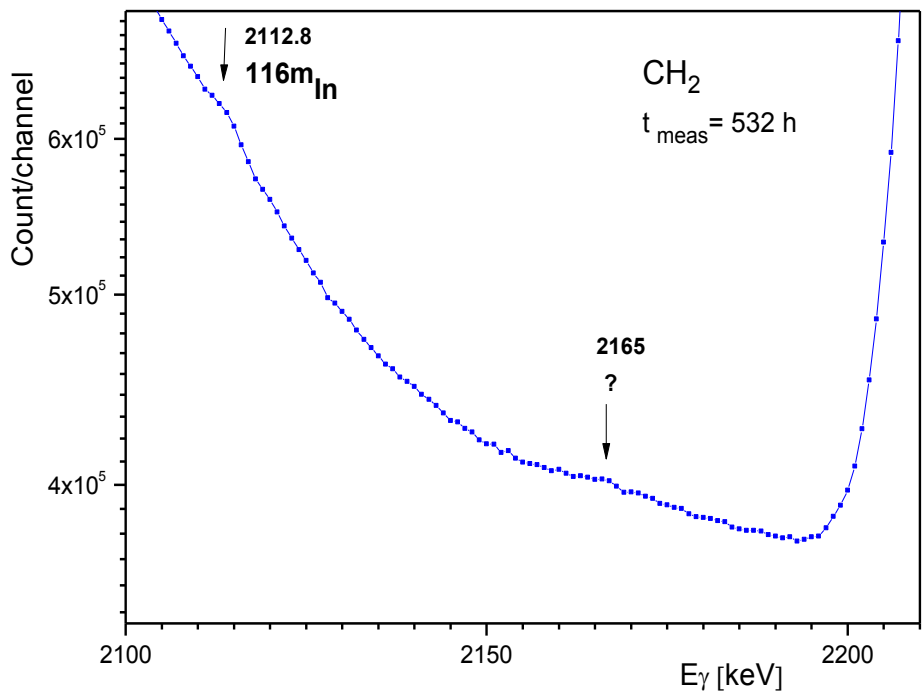

*Fig. 7. The part of the amplitude spectrum.*

The peak at 2223 keV contains $2.06 \cdot 10^8$ counts. In the energy interval 2100- 2210 keV we found only two peaks at 2112 and 2165 keV. The peak 2112 keV comes from $^{116m}$In decay. The line 2165 keV could be assigned to $^{56}$Fe but latter's yield (0.69%) is too small to be observed if to compare it with another observed $^{56}$Fe lines. The observed intensity in this peak is 6 times larger than anticipated from the above comparison. This peak has been observed in the previous preliminary measurement [29] (but not mentioned in the text) and yielded the cross section value of the assumed transition from hypothetical singlet state to be 46 ± 25 µb. The reported new result for this transition is 17± 6 µb which does not contradict to the preliminary result and needs additional measurements.

One more unrecognized peak is located at 453 keV, it was not found in the background spectrum. The assumed accompanying transition at 1770 keV is close to 1778.9 keV from $^{28}$Al, and is possibly not seen at this neighborhood.

The upper boundary of the cross section was estimated according to formula

$$\sigma_{n,2\gamma} \leq \frac{3\sqrt{S}}{S(2223)} \frac{\varepsilon(E_{\gamma 0})}{\varepsilon(E_{\gamma 1})} \sigma_{n,\gamma 0}$$

where $S$ is the sum in the energy range of the experimental peak width, $\sigma_{n,\gamma 0}$ is the cross section of the main transition 2223 keV, $\varepsilon(E_{\gamma 0})$ and $\varepsilon(E_{\gamma 1})$ are the detector efficiencies at the corresponding energies.

Table 3. The upper limit for $\sigma_{n,2\gamma}$.

| $E_\gamma$, keV | $\sigma_{n,2\gamma}$, µb |
|---|---|
| 2100 - 2130 | < 15 |
| 2131 – 2160 | < 12 |
| 2170 - 2200 | < 13 |

An upper limit at 3 standard deviations confidence level can be placed as shown in Table 3.

6. Conclusions.

We obtained new value for the upper limit for the two-step gamma-transition after slow neutron capture by protons $\sigma_{n,2\gamma}$ which is four times lower then in the preceding measurement [29]. This limit is lower than prediction [17]. The peak at the energy 2165 keV needs further investigation. Further search for the singlet deuteron on stationary reactor with improved techniques (e.g. suppressing of the Compton background) and statistics may be interesting.